\documentclass[aps,pra,floats,groupedaddress,showpacs, sort&compress, merge, superscriptaddress, smallextended]{revtex4}%
\usepackage{amsmath}
\usepackage{amsfonts}
\usepackage{amssymb}
\usepackage{graphicx}
\usepackage{dcolumn}
\usepackage{natbib}
\usepackage{color}
\usepackage{slashbox}
\usepackage{hyperref}

\newcommand{\la}{\langle}

\newcommand{\ra}{\rangle}

\newcommand{\executeiffilenewer}[3]{%
\ifnum\pdfstrcmp{\pdffilemoddate{#1}}%
{\pdffilemoddate{#2}}>0%
{\immediate\write18{#3}}\fi%
}

\begin{document}
\title{On the connection between quantum nonlocality and phase sensitivity of two-mode entangled Fock state superpositions}

\author{Kaushik P. Seshadreesan}
\email{ksesha1@lsu.edu}
\affiliation{Hearne Institute for Theoretical Physics and Department of Physics and Astronomy, Louisiana State University, Baton Rouge, LA 70803, USA}
\author{Christoph F. Wildfeuer}
\email{christoph.wildfeuer@gmail.com}
\affiliation{Institute of Mathematics and Natural Sciences, University of Applied Sciences and Arts Northwestern Switzerland, Bahnhofstrasse 6, 5210 Windisch, Switzerland}
\author{Moochan B. Kim}
\affiliation{Institute for Quantum Science and Engineering and Department of Physics and Astronomy, Texas A\&M University, College Station, TX 77843}
\author{Hwang Lee}
\affiliation{Hearne Institute for Theoretical Physics and Department of Physics and Astronomy, Louisiana State University, Baton Rouge, LA 70803, USA}
\author{Jonathan P. Dowling}
\affiliation{Hearne Institute for Theoretical Physics and Department of Physics and Astronomy, Louisiana State University, Baton Rouge, LA 70803, USA}
\date{\today}

\begin{abstract}
In two-mode interferometry, for a given total photon number $N$, entangled Fock state superpositions of the form $(|N-m\ra_a|m\ra_b+e^{i (N-2m)\phi}|m\ra_a|N-m\ra_b)/\sqrt{2}$ have been considered for phase estimation. Indeed all such states are maximally mode-entangled and violate a Clauser-Horne-Shimony-Holt (CHSH) inequality. However, they differ in their optimal phase estimation capabilities as given by their quantum Fisher informations. The quantum Fisher information is the largest for the $N00N$ state $(|N\ra_a|0\ra_b+e^{i N\phi}|0\ra_a|N\ra_b)/\sqrt{2}$ and decreases for the other states with decreasing photon number difference between the two modes. We ask the question whether for any particular Clauser-Horne (CH) (or CHSH) inequality, the maximal values of the CH (or the CHSH) functional for the states of the above type follow the same trend as their quantum Fisher informations, while also violating the classical bound whenever the states are capable of sub-shot-noise phase estimation, so that the violation can be used to quantify sub-shot-noise sensitivity. We explore CH and CHSH inequalities in a homodyne setup. Our results show that the amount of violation in those nonlocality tests may not be used to quantify sub-shot-noise sensitivity of the above states. 
\end{abstract}


\maketitle

\section{Introduction}
\label{intro}
Entanglement lies at the heart of various applications of quantum information processing, such as quantum computation, quantum key distribution, and quantum metrology~\cite{HHHH09}. In optical quantum information processing where interferometry is a widely used tool, the entanglement between the interfering modes is of high pertinence~\cite{Wei_12}. For example, in two-mode optical metrology~\cite{DJK15, *DS15, *jonmetrology}, numerous mode-entangled states of both definite and indefinite photon numbers have been proposed that achieve phase sensitivities beyond the shot-noise limit---also known as supersensitivity~\cite{Birrittella_12, *Joo_11, *kaushik1, *onoscheme, *petr, *pezzismerzi2, *uys, *holbur}. In the definite photon number case, the $N$-photon $N00N$ state $1/\sqrt{2}(|N\ra_{a}|0\ra_{b}+|0\ra_{a}|N\ra_{b})$ (where $a$, $b$ denote the modes) achieves the so-called Heisenberg limit $\delta\phi=1/N$~\cite{jonmetrology}, providing a quadratic enhancement over the shot-noise limit of $1/\sqrt{N}$. (Note that $\delta\phi$ denotes the uncertainty in the phase estimate. The smaller the value of $\delta\phi$, the better the precision.) Further, $N$-photon states of the form
\begin{align}
\label{mnm}
(N-m)::m\equiv\frac{1}{\sqrt{2}}(|N-m\ra_a|m\ra_b+e^{i (N-2m)\phi}|m\ra_a|N-m\ra_b),\nonumber\\
0\leq m< N/2, \  \{N,m\}\in \mathbb{Z}_{\geq 0},
\end{align}
which include the $N00N$ state ($m=0$), have been studied in the context of phase estimation in the presence of photon loss and have been shown to be potentially more robust than the $N00N$ states under such conditions~\cite{RoyBardhan_13, *Jiang1, *Huver1}. The role of entanglement in quantum enhanced phase estimation has been investigated extensively. It has been established that in two-mode linear interferometry entanglement in the probe photons is a necessary condition (although not sufficient) for achieving supersensitivity~\cite{Hyllus1, *pezzismerzi3, *glm}. 

Entanglement is intriguingly related to quantum nonlocality. Quantum nonlocality refers to the incompatibility of quantum mechanics with local hidden-variable theories, and is revealed by entangled quantum states via the violation of Bell's inequalities~\cite{Genovese_05, *Bellmain}. All entangled pure states of a multipartite system with multilevels are known to violate a Bell's inequality, a result known as Gisin's theorem~\cite{Sixia1, *Gisin_91, *Werner1}. The initial attempts at tests of Bell's inequalities were devised for, and performed on spin systems or on the polarization degree of freedom of photons. In 1998, Gilchrist {\it et al.}, showed that Bell-type quantum nonlocality tests could also be performed on optical quantum states entangled over spatial modes~\cite{Gilchrist_98}.  Using a continuous phase quadrature measurement, based on balanced homodyne detection, they showed that a pair-coherent state could be used to demonstrate Bell-type quantum nonlocality. Later, Munro considered such a measurement, and derived the optimal states that exhibited maximal Bell-inequality violations~\cite{Munro_99}. In 1999, Banaszek and W\'{o}dkiewicz introduced yet another approach to perform quantum nonlocality tests on mode-entangled states with an unbalanced homodyne-based setup followed by suitable photon number or parity measurements. They showed that the state $1/\sqrt{2}(|1\ra_{a}|0\ra_{b}+|0\ra_{a}|1\ra_{b})$ violates the Bell-Clauser-Horne (CH)~\cite{CHineq} and Bell-Clauser-Horne-Shimony-Holt (CHSH)~\cite{CHSHineq} inequalities with such a scheme~\cite{Banaszek1}. Using this scheme, Wildfeuer {\it et al.} studied the violations exhibited by $N00N$ states of different photon numbers $N$, and found strong violations for $N>1$~\cite{Wildfeuer1}. Later Gerry {\it et al.}, also using the the same scheme, showed that entangled coherent states exhibit even stronger maximal violations of the above Bell's inequalities than the $N00N$ states~\cite{Gerry3}. 

In this paper, we investigate the Bell violations exhibited by the $(N-m)::m$ states given in Eq.~(\ref{mnm}) with respect to CH and CHSH inequalities based on balanced and unbalanced homodyne detection. Contrary to the usual motivation of using entangled quantum states to demonstrate nonlocality of quantum mechanics, our motivation is to investigate the relationship, if any, between the quantum nonlocality and the phase sensitivity of the $(N-m)::m$ states. In Section~\ref{mnmsec}, we elaborate on this motive. We discuss some general attributes of the $(N-m)::m$ states for phase estimation. Further, we hypothesize a general relationship between quantum nonlocality and phase sensitivity of mode entangled states and infer expected trends in the quantum nonlocality of the $(N-m)::m$ states with respect to some particular (a priori unknown) Bell's inequality. In section~\ref{bsetup}, we briefly describe the homodyne-based Bell-testing schemes of Gilchrist {\it et al.}~\cite{Gilchrist_98} and Banaszek and W\'{o}dkiewicz~\cite{Banaszek1} to study quantum nonlocality of mode-entangled states. In section \ref{belltests} we study the maximal violations of CH and CHSH inequalities by the $(N-m)::m$ states in both the homodyne-based setups. In section \ref{sum}, we discuss the validity of our hypothesis in light of the results of the homodyne-based Bell-test for the $(N-m)::m$ states and conclude with a brief summary.

\section{Motivation} 
\label{mnmsec}

The $(N-m)::m$ states of Eq.~(\ref{mnm}) were introduced as a generalization of the $N00N$ state with nonzero photons in both modes. Just like the $N00N$ state, they are all maximally entangled. The logarithmic negativity parameter~\cite{Plenio_05, *Vidal_02}, which is a well-known measure of bipartite entanglement, captures this fact, since it evaluates to be $\mathcal{\varepsilon}=\log2$ for all such states. (See Appendix A for the calculation. We consider natural logarithm. Note that the value of logarithmic negativity for the states equals $\log$ of the Schmidt rank~\cite{Mike_Ike_00} of the states.) Also, they all violate the CHSH inequality given by
\begin{align}
\big\vert\langle A(\alpha)&\otimes B(\beta)\rangle+\langle A(\alpha')\otimes B(\beta)\rangle+\langle A(\alpha)\otimes B(\beta')\rangle-\langle A(\alpha')\otimes B(\beta')\rangle\big\vert\leq 2,
\end{align}
maximally at the Tsirelson bound~\cite{Cirel80} of $2\sqrt{2}$ using observables
\begin{align}
A(\alpha)=\Gamma_z \cos\alpha+\Gamma_x\sin\alpha,\nonumber\\
B(\beta)=\Gamma_z \cos\beta+\Gamma_x\sin\beta,
\end{align}
where
\begin{align}
\Gamma_z=\sum_{k=0}^{N}(-1)^{k}|k\rangle\langle k|,\nonumber\\
\Gamma_x=\sum_{k=0}^{N}|k\rangle\langle N-k|.
\end{align}
The optimal measurement settings are $\alpha=0$, $\alpha'=\pi/2$, $\beta=-\beta'$ as determined by Gisin and Peres~\cite{GP92}. 

\begin{figure}[h]
\centering
\includegraphics[scale=0.7]{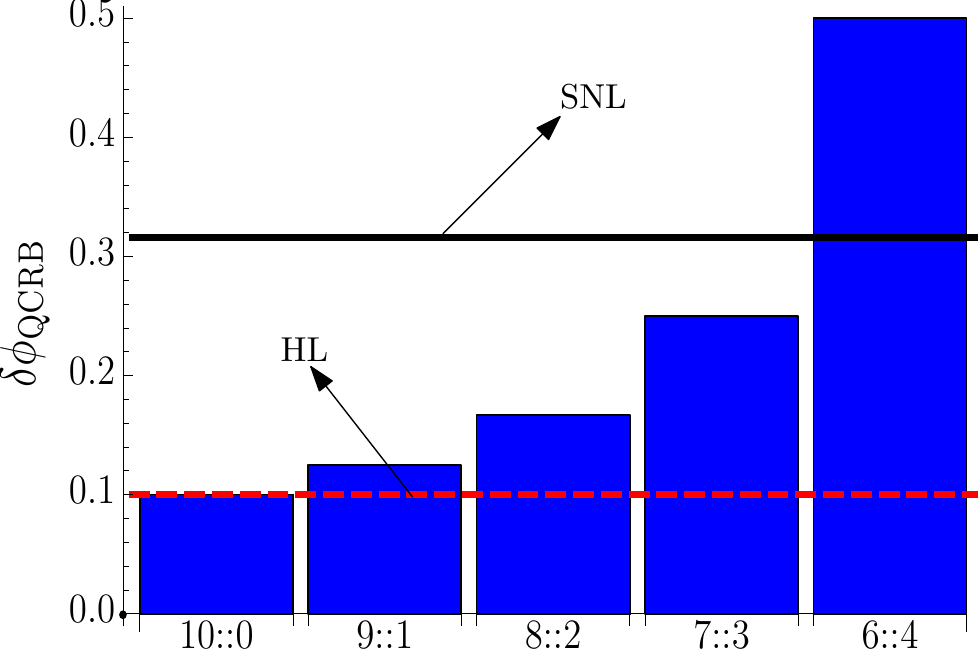}
\caption{(Color online) The QCRB on phase sensitivity of $(N-m)::m$ states of total photon number $N=10$, in the absence of photon losses. The phase sensitivity is described in terms of the minimum detectable phase change $\delta\phi$. The black (solid) line and the red (dashed) line denote the shot-noise (SNL, $1/\sqrt{N}$) and Heisenberg (HL, $1/N$) limits, respectively.}
\label{mnmqcrb}
\end{figure}

However, the states in Eq.~(\ref{mnm}) differ in their phase sensitivities in interferometric phase estimation. The quantum Cramer-Rao bound (QCRB)~\cite{QCRB2, *QCRB}, which is the inverse square-root of the quantum Fisher information (QFI), and determines the maximal phase sensitivity achievable with a state (independent of the detection scheme), captures this fact. The QCRB of a $(N-m)::m$ state under ideal lossless conditions evaluates to be
\begin{equation}
\label{qcrbmnm}
\delta\phi_{\rm QCRB}=\frac{1}{\sqrt{F_Q}}=\frac{1}{N-2m},
\end{equation}
where $F_Q$ denotes the QFI. (Note that the larger the QFI, the smaller the $\delta\phi_{\rm QCRB}$, which in turn implies better phase sensitivity.) Since the shot-noise limit for a state with total photon number $N$ is given by $\delta\phi=1/\sqrt{N}$, among the states in Eq.~(\ref{mnm}) for a total photon number $N$, only those that satisfy $m<(N-\sqrt{N})/2$ are capable of achieving supersensitivity. Fig.~(\ref{mnmqcrb}) shows a plot of the QCRB of the different states of the form in Eq.~(\ref{mnm}) corresponding to $N=10$. Evidently, the states $10::0$, $9::1$, $8::2$ and $7::3$ are capable of achieving phase supersensitivity.

While the $N00N$ states are capable of supersensitive phase estimation under lossless conditions, they decohere under loss of even a single photon, and with that drops their performance. Whereas, some of the $(N-m)::m$ states, although not optimally sensitive under ideal conditions, nevertheless perform better than the corresponding $N00N$ state in the presence of decoherence. Huver {\it et al.} showed that a $(N-m)::m$ state is immune to the loss of up to $m$ photons, and hence, can provide quantum-enhanced phase sensitivity under tolerable amounts of photon loss~\cite{Huver1}. Jiang {\it et al.} investigated the phase sensitivity and visibility of interference fringes of the states in Eq.~(\ref{mnm}) with a detection scheme based on photon number parity measurement, and prescribed the optimal choice of $N$ and $m$ for any given condition of photon loss~\cite{Jiang1}. Roy-Bardhan {\it et al.} showed that the optimal $(N-m)::m$ state in the presence of photon loss performs just as well as the corresponding $N00N$ state when dephasing noise is also present, while offering a better phase sensitivity than the latter given the photon loss~\cite{RoyBardhan_13}. 

\subsection*{The connection between quantum nonlocality and phase sensitivity: A Hypothesis}
\label{hypoes}

As mentioned above, the $(N-m)::m$ states, although all maximally entangled and nonlocal with respect to a CHSH inequality, differ in their phase sensitivities. We ask the question whether for any particular CHSH (or CH) inequality, the maximal values of the CHSH (or the CH) functional for the $(N-m)::m$ states follow the same trend as their QFIs, while also violating the classical bound whenever the states are capable of supersensitivity.  {\it We hypothesize quantum nonlocality (with respect to some particular a priori unknown CHSH or CH inequality) as an intrinsic resource for quantum phase estimation}. Based on the QCRB of $(N-m)::m$ states, we then anticipate the following trend in the maximal values of such a Bell's inequality functional for the different $(N-m)::m$ states:
\begin{enumerate}
\item{When a $(N-m)::m$ state is capable of supersensitivity, the maximal value of the functional must violate the Bell's inequality}
\item{Among the different $(N-m)::m$ states for a given total photon number $N$, the maximal value of the functional must be largest for the corresponding $N00N$ state, and should decrease with decrease in the photon number difference $N-2m$;}
\item{$(N-m)::m$ states of different total photon numbers $N$, but of the same photon number difference $N-2m$, must exhibit equal maximal values of the functional.}
\end{enumerate}
The statement 1 quintessentially captures the hypothesized connection between supersensitivity and quantum nonlocality. For a given total photon number $N$, the QCRB of Eq.~(\ref{qcrbmnm}) tells us that the optimal phase sensitivities of different $(N-m)::m$ states decreases when one moves from the $N::0$ state, i.e. the corresponding $N00N$ state, towards the separable state $|N/2\ra |N/2\ra$. That forms the basis of statement $2$. Also, we know that the optimal phase sensitivities of all $(N-m)::m$ states with the same photon number difference between the two modes, namely $N-2m$, are identical. That forms the basis of statement $3$.

In order to test the above hypothesis, we will consider CH and CHSH inequalities based on homodyne detection.

\section{Bell tests for mode-entangled states based on homodyne detection}
\label{bsetup}

First of all, we briefly review two different Bell-testing schemes for any state $|\psi\ra_{a,b}$ over two modes $\hat{a}$ and $\hat{b}$, proposed by Gilchrist {\it et al.}~\cite{Gilchrist_98} and Banaszek and W\'{o}dkiewicz~\cite{Banaszek1}, based on balanced and unbalanced homodyning, respectively.

\subsection{Balanced homodyning}
\label{balancedhom}

\begin{figure}[h]
\centering
\includegraphics[scale=0.5]{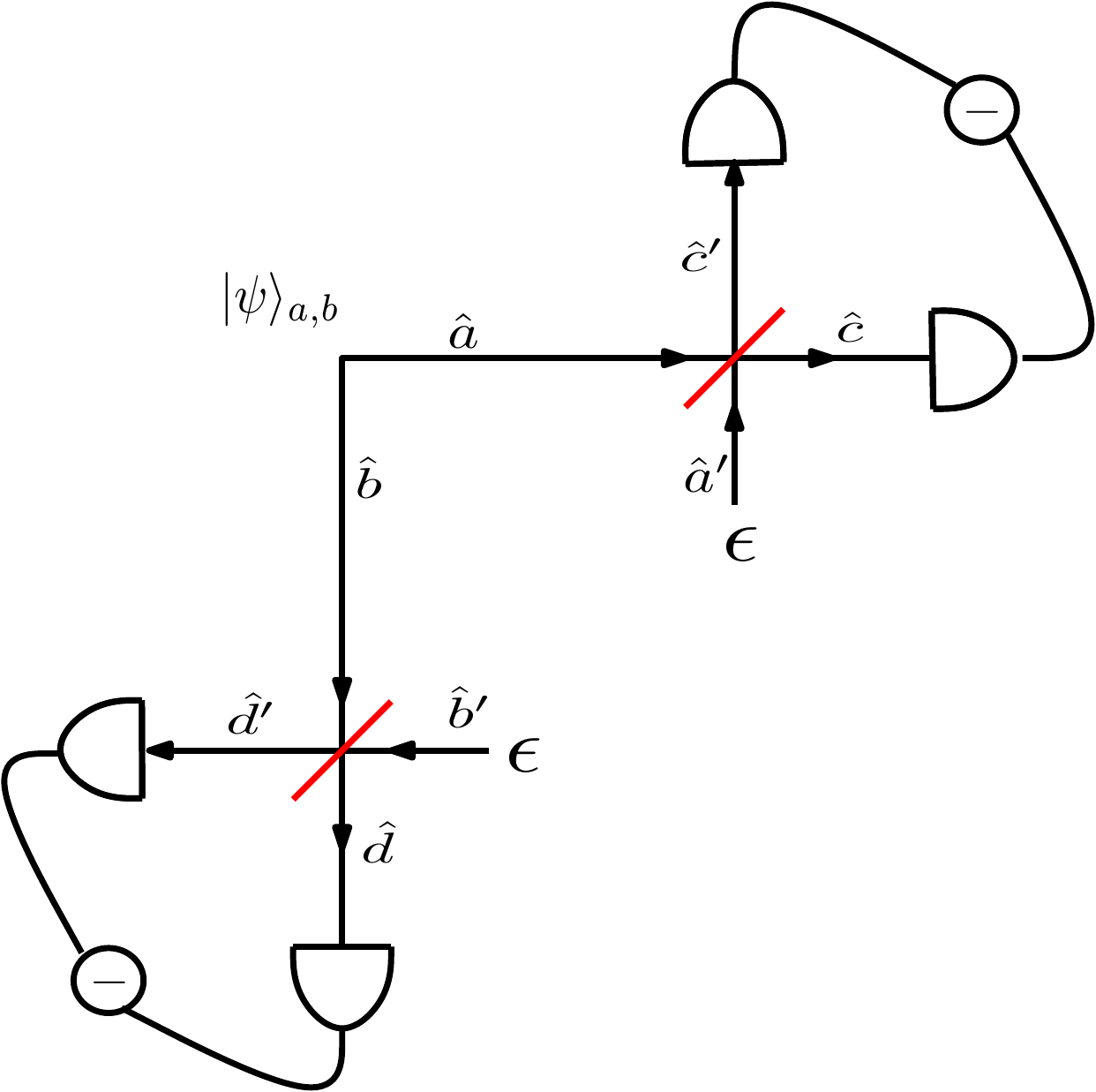}
\caption{(Color online) A schematic diagram of the balanced homodyne-based Bell-testing technique for a mode-entangled state $|\psi\rangle_{a,b}$. Strong classical local oscillator fields $\epsilon$ are mixed on to each of the two modes $\hat{a}$ and $\hat{b}$ through 50:50 beam splitters. The difference photocurrent in the output modes is then measured to determine the quadrature phase amplitudes.}
\label{bal}
\end{figure}

Consider the scheme described in Fig.~\ref{bal}. A strong local oscillator field $\epsilon$ is mixed on to each of the two modes $\hat{a}$ and $\hat{b}$ of the state $|\psi\ra_{a,b}$ through a 50:50 beam splitter, resulting in new field modes $\hat{c}=\left[\hat{a}+\epsilon\exp(i\theta)\right]/\sqrt{2}$, $\hat{c}'=\left[\hat{a}-\epsilon\exp(i\theta)\right]/\sqrt{2}$ and $\hat{d}$, $\hat{d}'$ likewise, respectively. The photocurrent differences $\hat{c}^\dagger\hat{c}-\hat{c}'^\dagger\hat{c}'$ and $\hat{d}^\dagger\hat{d}-\hat{d}'^\dagger\hat{d}'$, when measured, yield the quadrature phase amplitudes $\hat{X}_{\theta (\theta')}^{a}$ and $\hat{X}_{\varphi(\varphi')}^{b}$, given by
\begin{align}
\label{quadmeas}
\hat{X}_{\theta}^{a}=\hat{a}e^{-i\theta}+\hat{a}^\dagger e^{i\theta},\nonumber\\
\hat{X}_{\varphi}^{b}=\hat{b}e^{-i\varphi}+\hat{b}^\dagger e^{i\varphi},
\end{align}
(up to a factor of $\epsilon$). Here, $\theta$ and $\varphi$ act as control variables for the choice of observables to be measured on each half of the original two-mode state. Since we deal with bosonic modes, when the angle $\theta$ (or $\varphi$) is chosen to be 0 and $\pi/2$, these measurements correspond to measuring the position and momentum quadratures of the mode, respectively. The construction of Bell's inequalities requires the measurement of at least two non-commuting observables on each mode. For example, the CH and CHSH inequalities require the measurement of precisely two non-commuting observables on each mode. Therefore, by choosing distinct values for the local oscillator phases $\theta$, $\theta'$ (and $\varphi$, $\varphi'$), these Bell's inequalities can be tested using the balanced homodyne-based scheme.

The outcomes of the quadrature phase amplitude measurement of the above type on the two modes are continuous variables, which we denote by $x_1$ and $x_2$, respectively. In the joint measurement of observables $\hat{X}_{\theta}^{a}$ and $\hat{X}_{\varphi}^{b}$ on a state $|\psi\ra_{a,b}$, the probability of obtaining results $x_1$, $x_2$ is given by
 \begin{equation}
 \label{balhomprob}
 P_{x_1x_2}(\theta,\varphi)=|\la x_1,\ x_2|\psi_{a,b}\ra|^2,
 \end{equation}
where $|x_1\ra$ and $|x_2\ra$ are position eigenstates of the harmonic oscillator hamiltonian
\begin{equation}
\label{hermite}
\la x_i | n\ra=\frac{1}{\sqrt{2^n n! \sqrt{\pi}}}e^{-i n\alpha}e^{-x_i^2/2}H_n(x_i),
\end{equation}
$H_n(x_i)$ being the $n^{\rm th}$ Hermite polynomial, and $\alpha$ the phase of the local oscillator ($\theta$ for $x_1$ and $\varphi$ for $x_2$). For the $(N-m)::m$ state, $P_{x_1x_2}$ is found to be
\begin{align}
\label{mnmpx1x2}
&P_{x_1x_2}(\theta,\varphi)\nonumber\\
&=\frac{1}{2\pi}\frac{e^{-(x_1^2+x_2^2)}}{2^N (N-m)! m!}\left\vert e^{-i((N-m)\theta+m\varphi)}H_{N-m}(x_1)H_m(x_2)+e^{-i(m\theta+(N-m)\varphi-(N-2m)\phi)}H_m(x_1)H_{N-m}(x_2)\right\vert^2.
\end{align}

The CH and CHSH inequalities also require binary results at each mode. Thus, in order to test these inequalities, the continuous outcomes $x_1$ and $x_2$ are classified into two bins: 1 if $x_i\geq 0$ and 0 if $x_i<0$. The probability of obtaining different combinations of binary results at each mode, as well as marginal probabilities for different outcomes in each individual mode, can now be calculated by integrating $P_{x_1x_2}$ over suitable integration limits for $x_1$ and $x_2$. For instance, the probability of obtaining a result of  1 in both the modes, and the marginal probability of obtaining 1 in mode $\hat{a}$, and 1 in mode $\hat{b}$ are given by
\begin{align}
\label{11prob}
P_{11}(\theta,\varphi)=\int_{0}^{\infty}dx_1\int_{0}^{\infty}dx_2 P_{x_1x_2}(\theta,\varphi),\nonumber\\
P_{1a}(\theta)=\int_{0}^{\infty}dx_1 \int_{-\infty}^{\infty}dx_2 P_{x_1x_2}(\theta,\varphi),\nonumber\\
P_{1b}(\varphi)=\int_{-\infty}^{\infty}dx_1 \int_{0}^{\infty}dx_2 P_{x_1x_2}(\theta,\varphi).
\end{align}
Also, the correlation function $E(\theta,\varphi)\equiv\la\hat{X}_{\theta}^{a}\hat{X}_{\varphi}^{b}\ra$ can be calculated as
\begin{equation}
\label{Ecorrbal}
E(\theta,\varphi)=\int_{0}^{\infty}\int_{0}^{\infty}dx_1dx_2 \ {\rm sgn}(x_1x_2) P_{x_1x_2}(\theta,\varphi),
\end{equation}
where $\rm{sgn}(x)$ is the sign function, which is defined to be $1$ when the argument $x$ is greater than or equal to zero, and $-1$ otherwise.

\subsection{Unbalanced homodyning}
\label{unbalhom}

\begin{figure}[h]
\centering
\includegraphics[scale=0.5]{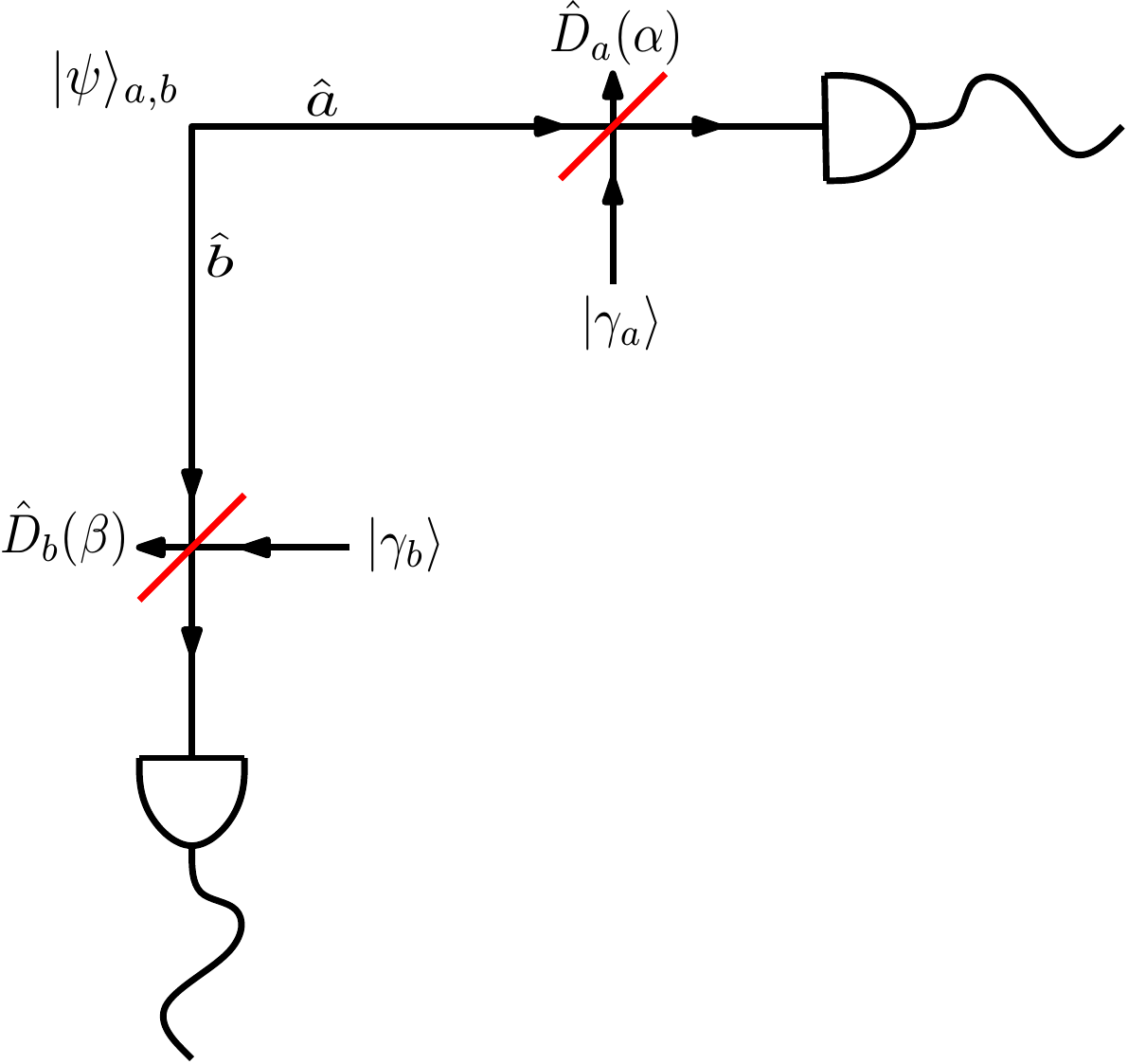}
\caption{(Color online) A schematic diagram of the unbalanced homodyne-based Bell-testing technique for a mode-entangled state $|\psi\rangle_{a,b}$. A strong local oscillator is mixed on to each of the two modes $\hat{a}$ and $\hat{b}$ through highly transmittive beam splitters (transmittivity $T\approx 1$), resulting in a displacement operation. The modes are subsequently detected using photon number or parity measurements.}
\label{unbal}
\end{figure}

This scheme is shown in Fig.~\ref{unbal}. In this case, strong coherent states $|\gamma_{a}\ra$ and $|\gamma_b\ra$ ($|\gamma_{a(b)}|\rightarrow\infty$) from a shared local oscillator, are mixed on the modes $\hat{a}$ and $\hat{b}$ of the state $|\psi\ra_{a,b}$ through highly transmittive beam splitters (the transmittivity $T\rightarrow 1$). This results in the beam splitters acting as effective displacement operators $\hat{D}(\gamma_{a(b)}\sqrt{1-T})$ on the modes $\hat{a}$ and $\hat{b}$, where $\hat{D}(\lambda)={\rm exp}(\lambda\hat{a}^{\dagger}-\lambda^{*}\hat{a})$~\cite{gerryknight}. The complex parameters $\alpha=\gamma_{a}\sqrt{1-T}$ and $\beta=\gamma_{b}\sqrt{1-T}$ act as control variables for measurements on the two modes analogous to the choice of the local oscillator phases in the balanced homodyning scheme described earlier.The two displaced modes that result from unbalanced homodyning are subsequently measured. We consider two different types of detectors to perform the measurements: a) on-off photodetectors and b) photon number parity detectors. For distinct values of the complex parameter in each mode, e.g., $\alpha$, $\alpha'$ in mode $\hat{a}$ (or $\beta$, $\beta'$ in mode $\hat{b}$), the observables corresponding to the action of these detectors on the displaced modes turn out to be non-commuting, thereby allowing to test the CH and CHSH inequalities using the scheme.

An on-off photodetector distinguishes between the events of `no incident photons' and `one or more incident photons'. Assuming lossless detectors, the positive operator valued measure (POVM) corresponding to such an operation following the unbalanced homodyning of Fig.~\ref{unbal}, e.g.~in mode $\hat{a}$, can be written as $\hat{Q}(\alpha)+\hat{P}(\alpha)=\hat{I}$, where
\begin{align}
\label{chpovm}
&\hat{Q}(\alpha)=\hat{D}(\alpha)|0\ra\la0|\hat{D}^{\dagger}(\alpha),\nonumber\\
&\hat{P}(\alpha)=\hat{D}(\alpha)\displaystyle\sum_{n=1}^{\infty}|n\ra\la n|\hat{D}^{\dagger}(\alpha).
\end{align}
For a normalized two-mode state $|\psi\rangle_{a,b}$, the expectation value $P_a(\alpha)=~\langle\psi|\hat{P}(\alpha)\otimes \hat{I} |\psi\rangle$ gives the probability of registering a single detector click in mode $\hat{a}$ (where the mode labels $a,\ b$ have been suppressed for convenience). Likewise, the expectation value $P_{ab}(\alpha,\beta)=~\langle\psi|\hat{P}_a(\alpha)\otimes\hat{P}_b(\beta)|\psi\rangle$ gives the probability of observing correlated detector click events in the two modes. The functions $P$ and $Q$ here are analogous to the probabilities of outcomes 1 and 0 in the balanced homodyning-based setup, respectively. Due to the completeness relation $\hat{Q}(\alpha)+\hat{P}(\alpha)=\hat{I}$, the single and correlated detector-click probabilities can be written in terms of the no-click probabilities $Q_a(\alpha)=~\langle\psi|\hat{Q}(\alpha)\otimes \hat{I} |\psi\rangle$, $Q_b(\beta)=~\langle\psi|\hat{I}\otimes \hat{Q}(\beta) |\psi\rangle$, and $Q_{ab}(\alpha,\beta)=~\langle\psi|\hat{Q}_a(\alpha)\otimes\hat{Q}_b(\beta)|\psi\rangle$, as
\begin{align}
\label{ptoq}
&P_a(\alpha)=1-Q_a(\alpha),\nonumber\\
&P_b(\beta)=1-Q_b(\beta),\nonumber\\
&P_{ab}(\alpha,\beta)=1-Q_a(\alpha)-Q_b(\beta)+Q_{ab}(\alpha,\beta).
\end{align}
The correlated no-click probability $Q_{ab}(\alpha,\beta)$ is related to the two-mode $Q$ function of the state $|\psi\rangle_{ab}$ (up to a factor of $1/\pi^2$), since $Q_{ab}(\alpha,\beta)=~\langle\psi|~(|\alpha\rangle\langle\alpha|\otimes|\beta\rangle\langle\beta|)~|\psi\rangle=|~\langle\alpha,\beta|\psi\rangle~|^2$. It is for this reason that Bell tests based on the unbalanced homodyning technique are sometimes referred to as tests for nonlocality in phase space. For the $(N-m)::m$ state, the functions $Q_a(\alpha)$, $Q_b(\beta)$ and $Q_{ab}(\alpha, \beta)$ are found to be
\begin{align}
\label{qfunc2}
&Q_a(\alpha;N,m)=\frac{1}{2}e^{-|\alpha|^2}\left(\frac{|\alpha|^{2(N-m)}}{(N-m)!}+\frac{|\alpha|^{2m}}{m!}\right)\nonumber\\
&Q_b(\beta;N,m)=\frac{1}{2}e^{-|\beta|^2}\left(\frac{|\beta|^{2(N-m)}}{(N-m)!}+\frac{|\beta|^{2m}}{m!}\right)\nonumber\\
&Q_{ab}(\alpha,\beta;N,m)=\frac{1}{2(N-m)!m!}e^{-(|\alpha|^2+|\beta|^2)}\nonumber\\
&\times\bigg[|\alpha^{N-m}\beta^{m}|^2+|\alpha^{m}\beta^{N-m}|^2+e^{i(N-2m)\phi}\left(\alpha\beta^*\right)^{N-m} \left(\alpha^*\beta\right)^{m}+e^{-i(N-2m)\phi}\left(\alpha\beta^*\right)^{m}\left(\alpha^*\beta\right)^{N-m}\bigg].
\end{align}
\par We now describe a photon number parity detector.
The photon number parity operator for a single mode is defined by
\begin{equation}
\label{pop}
\hat{\Pi}=\left(-1\right)^{\hat{n}}=\sum_{k=0}^{\infty}(|2k\ra\la2k|-|2k+1\ra\la2k+1|),
\end{equation}
where $\hat{n}$ is the number operator associated with the mode~\cite{gerry2}. It distinguishes between the even and odd photon number components of the state of the mode. The POVM for parity measurements in a single mode, following the unbalanced homodyning of Fig.~\ref{unbal}, can be written as $\hat{\Pi}(\alpha)=\hat{D}(\alpha)\hat{\Pi}\hat{D}^{\dagger}(\alpha)$. For a state $|\psi\rangle$, the joint measurement of the parity operators in the two modes leads to the correlation function 
\begin{eqnarray}
\label{Pi1}
\Pi_{ab}(\alpha, \beta)&=&~\langle\psi|\hat{\Pi}(\alpha)\otimes\hat{\Pi}(\beta)|\psi\rangle,\nonumber\\
&=&~\langle\psi|\hat{D}_a(\alpha)(-1)^{\hat{n}_a}\hat{D}_{a}^\dagger(\alpha)\otimes\hat{D}_b(\beta)(-1)^{\hat{n}_b}\hat{D}_{b}^\dagger(\beta)|\psi\rangle,
\end{eqnarray}
which is proportional to the two-mode Wigner function of the state (up to a factor of $4/\pi^2$). For the $(N-m)::m$ state, $\Pi_{ab}(\alpha, \beta)$ is found to be:
\begin{align}
\label{wfunc1}
&\Pi_{ab}(\alpha,\beta;N,m)=\frac{(-1)^N}{2}{\rm exp}(-2|\alpha|^2-2|\beta|^2)\nonumber\\
&\big\{L_{N-m}(4|\alpha|^2)L_{m}(4|\beta|^2)+L_{m}(4|\alpha|^2)L_{N-m}(4|\beta|^2)\nonumber\\
&+2{\rm Re}[{\rm exp}(i(N-2m) \phi)(\frac{\alpha}{\beta})^{N-2m} L_{m}^{N-2m}(4|\alpha|^2)L_{N-m}^{-(N-2m)}(4|\beta|^2)]\big\},
\end{align}
where $L_{i}^{j}$ are the associated Laguerre polynomials~\cite{Arfken_85}. (See Appendix B for the derivation.)

\section{Quantum nonlocality of the $(N-m)::m$ states based on the above homodyne Bell tests}
\label{belltests}

We now describe the testing of CH and CHSH inequalities for the $(N-m)::m$ states of Eq.~(\ref{mnm}) based on the two homodyne-based Bell-testing techniques described above.

\subsection{The Bell-Clauser-Horne test}
\label{ch}

The CH inequality is a constraint on classical correlations in the space of probability distributions. It can be constructed out of joint probabilities $P(x_1, x_2 | X_a(X_a'), X_b(X_b'))$ for obtaining outcomes $x_1$, $x_2\in \{0,1\}$, and the associated marginal probabilities, where $X_a$, $X_a'$ and $X_b$, $X_b'$ are pairs of non-commuting observables measured on modes $\hat{a}$ and $\hat{b}$, respectively. For the balanced homodyne-based setup, we construct the CH inequality with the probabilities $P_{11}$ of Eq.~(\ref{11prob}) corresponding to the the joint measurement of observables $\hat{X}_{\theta}^{a}$ ($\hat{X}_{\theta'}^{a}$) and $\hat{X}_{\varphi}^{b}$ ($\hat{X}_{\varphi'}^{b}$) on the $(N-m)::m$ state as $-1\leq \mathcal{CH}\leq 0$, where
\begin{eqnarray}
\label{chineqbal}
&\mathcal{CH}=P_{11}(\theta,\varphi)-P_{11}(\theta,\varphi')+P_{11}(\theta',\varphi)+P_{11}(\theta',\varphi')-P_{1a}(\theta')-P_{1b}(\varphi).&
\end{eqnarray}
We numerically extremize $\mathcal{CH}$ over the space of the parameters $\theta,\ \theta',\ \varphi,\ \varphi'$ for $(N-m)::m$ states up to $N=9$. The results are obtained using a Mathematica subroutine, which implements a random-search algorithm for global optimization. The extremal values are tabulated in Table~\ref{table1}. We find that none of the $(N-m)::m$ states violate either of the bounds of the inequality.

\begin{table}[ht]
\centering
\begin{tabular}{|c|c|c|c|c|c|c|}
	\hline
\backslashbox{N-m}{m} & $0$ & $1$ & $2$ & $3$ & $4$ & $5$ \\
\hline
$0$ && $\bf{-0.18}$ & $\bf{-0.50}$ & $\bf{-0.45}$ & $\bf{-0.50}$ & $\bf{-0.48}$\\
\hline
$1$ & $-0.82$ && $\bf{-0.34}$ & $\bf{-0.50}$ & $\bf{-0.49}$ & $\bf{-0.50}$\\
\hline
$2$ & $-0.50$ & $-0.66$ && $\bf{-0.26}$ & $\bf{-0.50}$ & $\bf{-0.47}$\\
\hline
$3$ & $-0.55$ & $-0.50$ & $-0.74$ & & $\bf{-0.32}$ & $\bf{-0.50}$\\
\hline
$4$ & $-0.50$ & $-0.51$ & $-0.50$ & $-0.68$ & & $\bf{-0.28}$\\
\hline
$5$ & $-0.52$ & $-0.50$ & $-0.53$ & $-0.50$ & $-0.72$ &\\
\hline
\end{tabular}
\caption{The extremal values of  $\mathcal{CH}$ for the $(N-m)::m$ states. Values in the lower triangle of the table are the minimum values for states $(N-m)::m$, and those in the upper triangle of the table (bold faced) are the maximum values for states $m::N-m$. We see that none of the states violate the CH inequality.} 
\label{table1} 
\end{table}

\begin{figure}[h]
\centering
\includegraphics[scale=0.85]{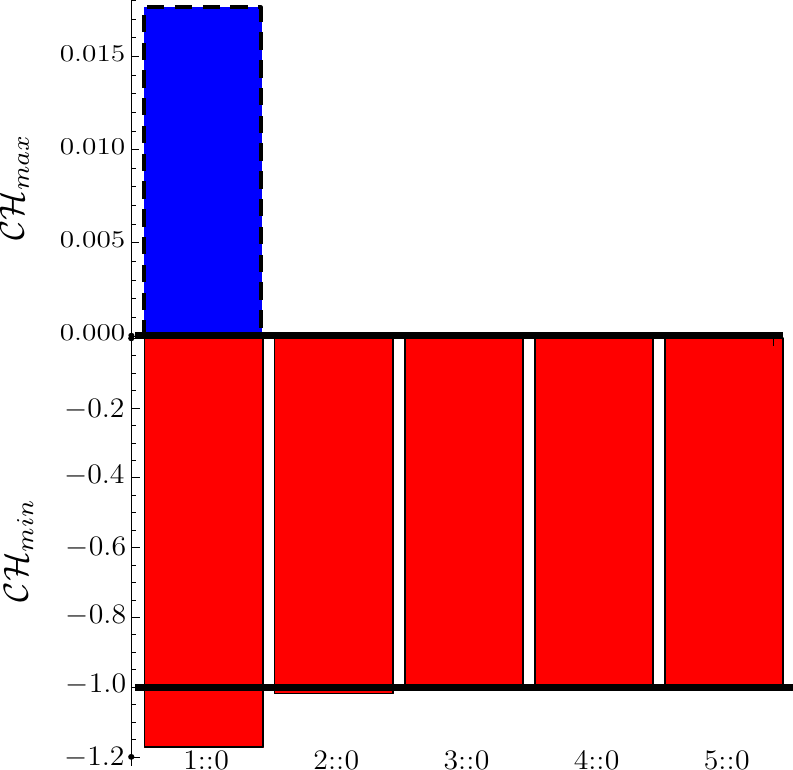}
\caption{(Color online) The maximum (blue, dashed borders) and minimum (red, solid borders) values of the $\mathcal{CH}$ functional for $N00N$ states of $N=1,\ 2, ...,\ 5$ photons. Lines $y=0$ and $y=-1$ represent the upper and lower bounds of the CH inequality. We find that states $1::0$ and $2::0$ violate the lower bound, and the state $1::0$ alone violates the upper bound of the inequality.}
\label{ch_noon}
\end{figure}

\begin{figure}[h]
\centering
\includegraphics[scale=0.85]{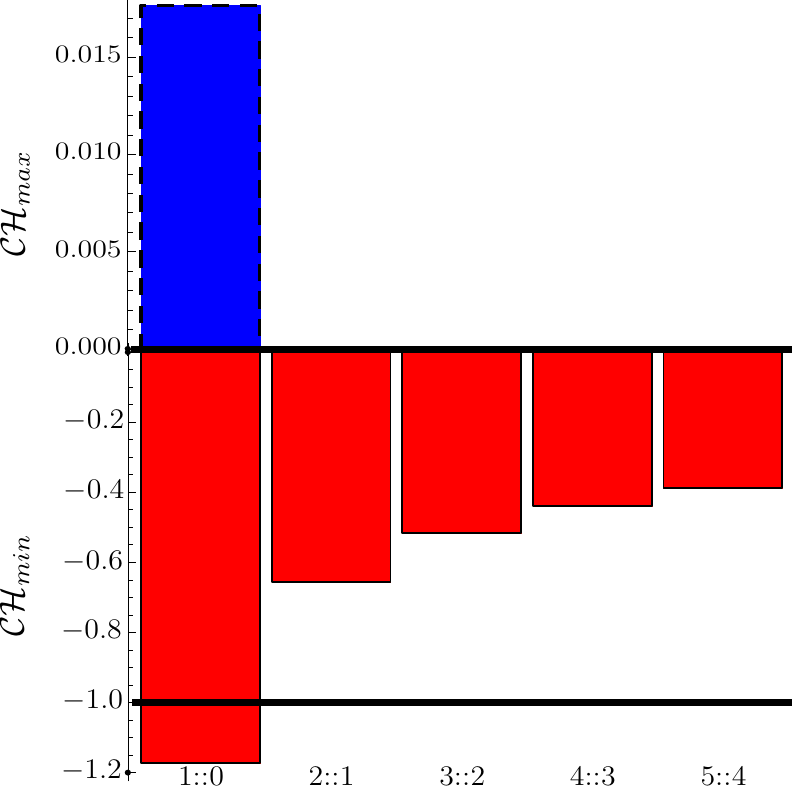}
\caption{(Color online) The maximum (blue, dashed borders) and minimum (red, solid borders) values of the $\mathcal{CH}$ functional for the $(N-m)::m$ states with photon number difference $N-2m=1$, for up to a total photon number of $N=9$. Lines $y=0$ and $y=-1$ represent the upper and lower bounds of the CH inequality. We find that none of the $(N-m)::m$ states, except the $1::0$ $N00N$ state violate either of the bounds of the inequality.}
\label{ch_m1}
\end{figure}

For the unbalanced homodyne-based set up, we construct the $\mathcal{CH}$ combination using the single and correlated detector-click probabilities of Eq.~(\ref{ptoq}) as 
\begin{eqnarray}
\label{chineq}
\mathcal{CH}=P_{ab}(\alpha,\beta)-P_{ab}(\alpha,\beta')+P_{ab}(\alpha',\beta)+P_{ab}(\alpha',\beta')-P_{a}(\alpha')-P_{b}(\beta).
\end{eqnarray}
We numerically extremize $\mathcal{CH}$ over the space of the complex parameters $\alpha$, $\alpha'$, $\beta$ and $\beta'$ for $(N-m)::m$ states up to $N=9$. (Note that we optimize over both the real and imaginary parts of these complex parameters). The results are obtained using a mathematica subroutine, which implements a random-search algorithm for global optimization. Fig.~\ref{ch_noon} shows plots of the maximum and minimum values of $\mathcal{CH}$ for $N00N$ states up to $N=5$. We see that the lower bound of $-1$ is violated by the $N=1,\ 2$ states. Wildfeuer {\it et al.} showed analytically that in fact all $N00N$ states corresponding to finitely large photon numbers $N$ violate the lower bound of the inequality, however, by increasingly smaller amounts for larger values of $N$~\cite{Wildfeuer1}. As for the maximum values, we find that the state corresponding to $N=1$ alone violates the upper bound of 0, while all other higher $N00N$ states optimally attain the bound value of 0. Fig.~\ref{ch_m1} shows plots of the maximum and minimum values of $\mathcal{CH}$, respectively, for $(N-m)::m$ states with $N-2m=1$, for up to a total photon number $N=9$. We find that none of the states, other than the $1::0$ $N00N$ state, violate either of the bounds of the CH inequality. Similar optimizations were carried out for $N-2m=2$ $(N-m)::m$ states. Other than the $N00N$ states, none of them were found to violate the inequality.

\subsection{The Bell-Clauser-Horne-Shimony-Holt Test}
\label{chsh}

The CHSH inequality is a constraint based on correlation functions $-1\leq\Pi_{ab}(X_a(X'_a),X_b(X'_b))\leq 1$, where $X_a$, $X_a'$ and $X_b$, $X_b'$ are pairs of non-commuting observables measured on modes $\hat{a}$ and $\hat{b}$, respectively. For the balanced homodyne-based setup, we construct the CHSH inequality with the correlation function $E$ of Eq.~(\ref{Ecorrbal}) corresponding to the joint measurement of observables $\hat{X}_{\theta}^{a}$ ($\hat{X}_{\theta'}^{a}$) and $\hat{X}_{\varphi}^{b}$ ($\hat{X}_{\varphi'}^{b}$) on the $(N-m)::m$ state as $-2\leq \mathcal{CHSH}\leq 2$, where
\begin{equation}
\label{chshineq}
\mathcal{CHSH}=E(\theta,\varphi)+E(\theta',\varphi)+E(\theta,\varphi')-E(\theta',\varphi').
\end{equation}
The value of $\mathcal{CHSH}$ for the $(N-m)::m$ states of up to $N=9$, extremized over the space of the parameters $\theta,\ \theta',\ \varphi,\ \varphi'$, are tabulated in Table~\ref{table2}. Once again, we find that none of the $(N-m)::m$ states violate any of the bounds of the inequality.

\begin{table}[ht]
\centering
\begin{tabular}{|c|c|c|c|c|c|c|}
	\hline
\backslashbox{N-m}{m} & $0$ & $1$ & $2$ & $3$ & $4$ & $5$ \\
\hline
$0$ && $\bf{1.27}$ & $\bf{0}$ & $\bf{0.21}$ & $\bf{0}$ & $\bf{0.10}$\\
\hline
$1$ & $-1.27$ && $\bf{0.64}$ & $\bf{0}$ & $\bf{0.05}$ & $\bf{0}$\\
\hline
$2$ & $0$ & $-0.64$ && $\bf{0.95}$ & $\bf{0}$ & $\bf{0.13}$\\
\hline
$3$ & $-0.21$ & $0$ & $-0.95$ & & $\bf{0.72}$ & $\bf{0}$\\
\hline
$4$ & $0$ & $-0.05$ & $0$ & $-0.72$ & & $\bf{0.90}$\\
\hline
$5$ & $-0.10$ & $0$ & $-0.13$ & $0$ & $-0.90$ &\\
\hline
\end{tabular}
\caption{The extremal values of  $\mathcal{CHSH}$ for the $(N-m)::m$ states. Values in the lower triangle of the table are the minimum values for states $(N-m)::m$, and the those in the upper triangle of the table (bold faced) are the maximum values for states $m::N-m$. We see that none of the states violate the CHSH inequality.} 
\label{table2} 
\end{table}

\begin{figure}[h]
\centering
\includegraphics[scale=0.85]{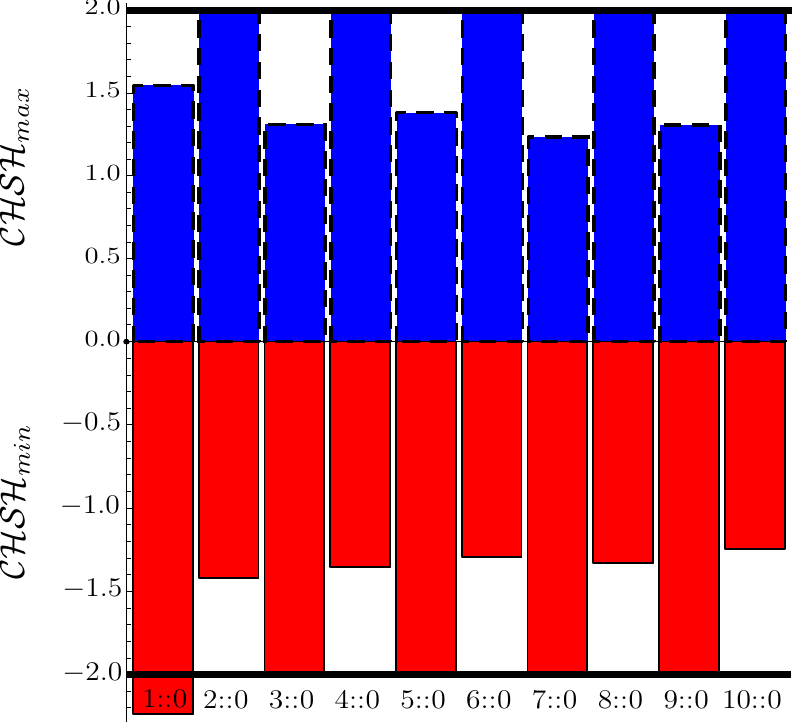}
\caption{(Color online) The maximum (blue, dashed borders) and minimum (red, solid borders) values of the $\mathcal{CHSH}$ functional for $N00N$ states of $N=1,\ 2, ...,\ 10$ photons. Lines $y=+2$ and $y=-2$ represent the upper and lower bounds of the CHSH inequality. We find that the $1::0$ state violates the lower bound, while none of the states violate the upper bound of the inequality.}
\label{chsh_1}
\end{figure}

\begin{figure}[h]
\centering
\includegraphics[scale=0.85]{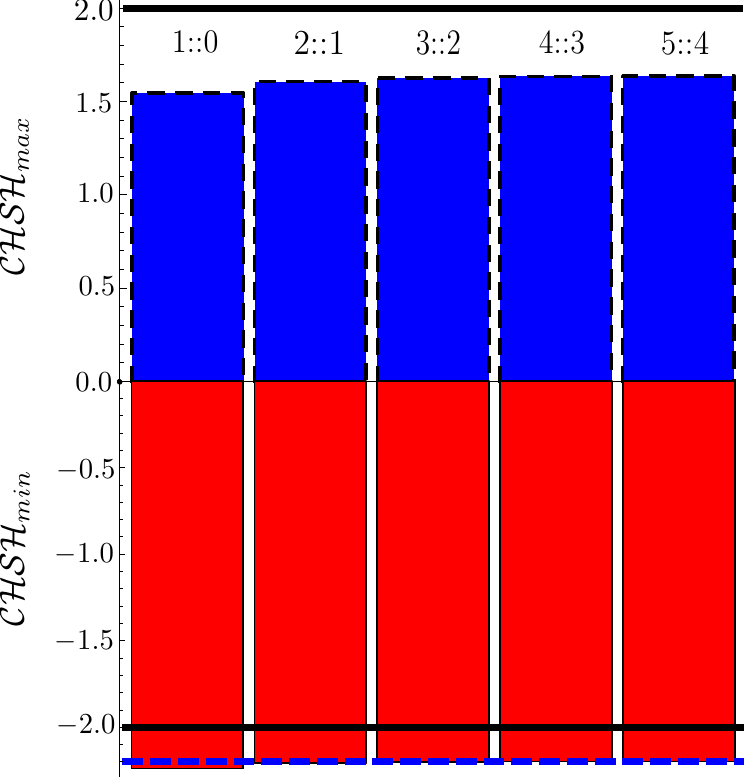}
\caption{(Color online) The maximum (blue, dashed borders) and minimum (red, solid borders) values of the $\mathcal{CHSH}$ functional for $(N-m)::m$ states with photon number difference $N-2m=1$, for up to a total photon number of $N=9$. Lines $y=+2$ and $y=-2$ represent the upper and lower bounds of the CHSH inequality. We find that all the states violate the lower bound of the inequality, reaching a value of $-2.2$, while none of the states violate the upper bound of the inequality.}
\label{chsh_2}
\end{figure}

For the unbalanced homodyne-based set up, we construct $\mathcal{CHSH}$ using the parity-based correlation function $\Pi_{ab}$ of Eq.~(\ref{Pi1}) as
\begin{equation}
\label{chshineq}
\mathcal{CHSH}=\Pi_{ab}(\alpha,\beta)+\Pi_{ab}(\alpha',\beta)+\Pi_{ab}(\alpha,\beta')-\Pi_{ab}(\alpha',\beta').
\end{equation}
We numerically extremize $\mathcal{CHSH}$ over the space of the complex parameters $\alpha$, $\alpha'$, $\beta$ and $\beta'$ for $(N-m)::m$ states up to $N=10$. Fig.~\ref{chsh_1} shows plots of the maximum and minimum values of $\mathcal{CHSH}$ for $N00N$ states for up to $N=10$. We see that the minimum value of the functional for the $1::0$ $N00N$ state alone violates the lower bound $-2$ of the inequality, a result already shown in Ref.~{\cite{Wildfeuer1}}. As for the upper bound, we find that none of the $N00N$ states violate the bound of $2$. Fig.~\ref{chsh_2} shows plots of the maximum and minimum values of $\mathcal{CHSH}$ for the $(N-m)::m$ states with photon number difference $N-2m=1$ and a total photon number $N=9$. Interestingly, we find that all $(N-m)::m$ states with $N-2m=1$ violate the lower bound of $-2$, reaching a minimum value of $\approx-2.2$. Needless to say, none of them violate the upper bound of $2$. Similar optimizations were carried out for $N-2m=2$, $N-2m=3$ $(N-m)::m$ states; none of them were found to violate either of the bounds of the inequality.

\section{Conclusion and Summary}
\label{sum}

Evidently, the results of Sec.~\ref{belltests} do not comply with the statement 1 of the hypothesis presented in Sec.~\ref{hypoes}. A majority of the $(N-m)::m$ states (the $N00N$ states and states with photon number difference $N-2m=1$ being the only exceptions) do not exhibit any violation whatsoever of the two Bell's inequalities in the considered balanced and unbalanced homodyne setups. Thus, the violations of CH and CHSH inequalities by the $(N-m)::m$ states in the considered homodyne-based setups do not support any connection between quantum nonlocality and the phase sensitivity of the states. We note that the quadratures measured in a homodyne detection scheme depend on the local oscillator coherent fields.  However, the QFI and hence the QCRB only depend on the photon numbers of the two mode quantum state. This supports the results that we find no simple relation between the CH and CHSH violations in homodyne detection schemes, and the phase sensitivity for $(N-m)::m$ states, respectively.

In summary, we studied the Bell-type quantum nonlocality exhibited by the two-mode entangled Fock state superpositions of Eq.~(\ref{mnm}), $(N-m)::m$. We considered the quantum nonlocality of the states with respect to Clauser-Horne and Clauser-Horne-Shimony-Holt inequalities in balanced and unbalanced homodyne detection schmes of Gilchrist {\it et al.}~\cite{Gilchrist_98} and Banaszek and W\'{o}dkiewicz~\cite{Banaszek1}. We made an attempt to identify the relationship between the quantum nonlocality and the phase sensitivity of such states. We found that the Bell tests performed using the said homodyne-based schemes do not support any connection between the two quantities for $(N-m)::m$ states of Eq.~(\ref{mnm}). Hence, the amount of Bell violation in a homodyne setup as considered here may not be used to quantify sub-shot-noise sensitivity. However, the connection of Bell violation and sub-shot-noise sensitivity in other quantum metrology schemes remains a topic of ongoing research.

\textbf{Acknowledgments.} JPD and KPS would like to acknowledge support from the Air Force Office of Scientific Research, the Army Research Office, the Defense Advanced Research Projects Agency, and the National Science Foundation. CFW would like to add the following acknowledgement: I first met Howard at QCMC 2002 in Boston. He told me that as part of his formal education he went to the south of Germany and attended High School for a couple of years. He enjoyed chatting in German with me. Howard also introduced me to JPD at this conference who later hired me as a post doc. I met Howard several times at various international conferences over the past years. He always gave me advice on career moves and research topics, even in the weeks before his surgery. His encouragement and generosity made a deep impression on me and I will never forget this extraordinary researcher and warm welcoming personality.

\appendix
\section*{Appendix A}
\setcounter{section}{1}
\setcounter{equation}{0}
Let $\rho$ be the density operator corresponding to the $(N-m)::m$ state of Eq.~({\ref{mnm}}), i.e.
\begin{align}
\label{densop}
\rho&=| (N-m)::m\rangle\langle (N-m)::m|\nonumber\\
&=\frac{1}{2}\bigg[|N-m,m\rangle\langle N-m,m|+|m,N-m\rangle\langle m,N-m|\nonumber\\
&+ e^{-i(N-2m) \phi}|N-m,m\rangle\langle m,N-m|+e^{i(N-2m) \phi}|m,N-m\rangle\langle N-m,m|\bigg].
\end{align}
The logarithmic negativity $\mathcal{\varepsilon}$ of the state can be calculated using the absolute sum of the negative eigenvalues $\mathcal{N}=|\sum_{i}\lambda_i|$, $\lambda_i < 0$ of the partial transpose of the density operator $\rho^{PT}$, as $\mathcal{\varepsilon}=\log{(1+2\mathcal{N})}$. The partial transpose of $\rho$ of Eq.~(\ref{densop}) is given by
\begin{align}
\label{ptrans}
&\rho^{PT}=| (N-m)::m\rangle\langle (N-m)::m|\nonumber\\
&=\frac{1}{2}\bigg[|N-m,m\rangle\langle N-m,m|+|m,N-m\rangle\langle m,N-m|\nonumber\\
&+ e^{-i(N-2m) \phi}|N-m,N-m\rangle\langle m,m|+e^{i(N-2m) \phi}|m,m\rangle\langle N-m,N-m|\bigg].
\end{align}
Diagonalizing the off-diagonal terms, $\rho^{PT}$ of Eq.~(\ref{ptrans}) can be equivalently written as
\begin{align}
\label{doffd}
\rho&=| (N-m)::m\rangle\langle (N-m)::m|\nonumber\\
&=\frac{1}{2}\bigg[|N-m,m\rangle\langle N-m,m|+|m,N-m\rangle\langle m,N-m|+|\varphi_1\rangle\langle\varphi_1|-|\varphi_2\rangle\langle\varphi_2|\bigg],
\end{align}
where $|\varphi_1\rangle$ and $|\varphi_2\rangle$ are normalized two-mode states given by
\begin{align}
\label{diag}
|\varphi_1\rangle=\frac{1}{\sqrt{2}}\left(e^{-i (N-m) \phi}|N-m,N-m\rangle+e^{-i m \phi}|m,m\rangle\right),\nonumber\\
|\varphi_2\rangle=\frac{1}{\sqrt{2}}\left(e^{-i (N-m) \phi}|N-m,N-m\rangle-e^{-i m\phi}|m,m\rangle\right).
\end{align} 
The eigenvalues of $\rho^{PT}$ are $\{1/2,\ 1/2,\ 1/2,\ -1/2\}$. We notice that they are independent of the values of $N$ and $m$. Thus, the logarithmic negativity of all $(N-m)::m$ states is $\log2$.
\\
\section*{Appendix B}
\setcounter{section}{2}
\setcounter{equation}{0}
The correlation function $\Pi(\alpha, \beta)$ of Eq.~(\ref{Pi1}), for an $(N-m)::m$ state of Eq.~(\ref{mnm}), is given by
\begin{align}
\label{appeq2}
&\la (N-m)::m|\hat{\Pi}(\alpha)\otimes\hat{\Pi}(\beta)|(N-m)::m\ra=\nonumber\\
&\frac{1}{2}\bigg[\la N-m|\hat{D}(\alpha)(-1)^{\hat{n}_a} \hat{D}(-\alpha)|N-m\ra\la m|\hat{D}(\beta)(-1)^{\hat{n}_b} \hat{D}(-\beta)|m\ra\nonumber\\
&+\la m|\hat{D}(\alpha)(-1)^{\hat{n}_a} \hat{D}(-\alpha)|m\ra\la N-m |\hat{D}(\beta)(-1)^{\hat{n}_b} \hat{D}(-\beta)|N-m\ra\nonumber\\
&+\{{\rm exp}(i(N-2m)\phi)\la N-m|\hat{D}(\alpha)(-1)^{\hat{n}_a} \hat{D}(-\alpha)|m\ra\times\la m |\hat{D}(\beta)(-1)^{\hat{n}_b} \hat{D}(-\beta)|N-m\ra+{\rm c.c.}\}\bigg],
\end{align}
where we have used the fact that $\hat{D}(\alpha)^{\dagger}=\hat{D}(-\alpha)$. Denoting displaced Fock states by $\hat{D}(\alpha)|n\ra=|\alpha,n\ra$, Eq.~(\ref{appeq2}) can be rewritten as
\begin{align}
\label{appeq3}
&\la (N-m)::m|\hat{\Pi}(\alpha)\otimes\hat{\Pi}(\beta)|(N-m)::m\ra=\nonumber\\
&\frac{1}{2}\bigg[\la -\alpha,N-m|(-1)^{\hat{n}_a}|-\alpha,N-m\ra\la -\beta, m|(-1)^{\hat{n}_b}|-\beta,m\ra\nonumber\\
&+\la -\alpha,m|(-1)^{\hat{n}_a}|-\alpha,m\ra\la -\beta, N-m|(-1)^{\hat{n}_b}|-\beta,N-m\ra\nonumber\\
&+2{\rm Re}\{{\rm exp}(i(N-2m)\phi)\la -\alpha,N-m|(-1)^{\hat{n}_a}|-\alpha,m\ra\la -\beta, m|(-1)^{\hat{n}_b}|-\beta,N-m\ra\}\bigg].
\end{align}
Using the number basis expansion of states of the form $|\alpha,n\ra$ as given in Ref. \cite{Knight90}, one can show that
\begin{equation}
\label{appeq4}
(-1)^{\hat{n}}|\alpha,N-m\ra=(-1)^{n}|-\alpha,n\ra.
\end{equation}
Therefore, Eq.~(\ref{appeq3}) can be written as:
\begin{align}
\label{appeq5}
&\la (N-m)::m|\hat{\Pi}(\alpha)\otimes\hat{\Pi}(\beta)|(N-m)::m\ra=\nonumber\\
&\frac{(-1)^{N}}{2}\bigg[\la -\alpha,N-m|\alpha,N-m\ra\la -\beta, m|\beta,m\ra+\la -\alpha,m|\alpha,m\ra\la -\beta, N-m|\beta,N-m\ra\nonumber\\
&+2{\rm Re}\{{\rm exp}(i(N-2m)\phi)\la -\alpha,N-m|\alpha,m\ra\la -\beta, m|\beta,N-m\ra\}\bigg].
\end{align}
The inner product of displaced Fock states is given by~\cite{Wunsche91}
\begin{align}
\label{appeq6}
\la-\alpha,N-m|\alpha,m\ra={\rm exp}(-2|\alpha|^2)&\sqrt{\frac{m!}{(N-m)!}}(2\alpha)^{N-2m}L_{N-m}^{N-2m}(4|\alpha|^2).
\end{align}
Using Eq.~(\ref{appeq6}) in Eq.~(\ref{appeq5}), the correlation function $\Pi(\alpha, \beta)$ for the $(N-m)::m$ state is found to be of the form given in Eq.~(\ref{wfunc1}). Further, the two-mode Wigner function of the state can be written as
\begin{equation}
\label{appeq1}
W(\alpha,\beta)=\frac{4}{\pi^2}\Pi(\alpha,\beta).
\end{equation}

\bibliographystyle{unsrt}
\bibliography{references_1}
\end{document}